\pdfoutput=1

\documentclass[11pt]{article}

\usepackage[preprint]{acl}

\usepackage{times}
\usepackage{latexsym}
\usepackage{multirow}
\usepackage{graphicx}
\usepackage[T1]{fontenc}

\usepackage[utf8]{inputenc}

\usepackage{microtype}
\usepackage{inconsolata}
\usepackage{booktabs}
\usepackage{graphicx}

\title{QUST\_NLP at SemEval-2025 Task 7: A Three-Stage Retrieval Framework for Monolingual and Crosslingual Fact-Checked Claim Retrieval}

\author{Youzheng Liu\textsuperscript{1}, Jiyan Liu\textsuperscript{1}, Xiaoman Xu\textsuperscript{1}, Taihang Wang\textsuperscript{1}\thanks{Corresponding author}, Yimin Wang\textsuperscript{2} \and Ye Jiang\textsuperscript{1}  \\
        College of Information Science and Technology\textsuperscript{1} \\ 
        College of Data Science\textsuperscript{2} \\
        Qingdao University of Science and Technology, China}

\begin{document}
\maketitle
\begin{abstract}


This paper describes the participation of QUST\_NLP in the SemEval-2025 Task 7. We propose a three-stage retrieval framework specifically designed for fact-checked claim retrieval. Initially, we evaluate the performance of several retrieval models and select the one that yields the best results for candidate retrieval. Next, we employ multiple re-ranking models to enhance the candidate results, with each model selecting the Top-10 outcomes. In the final stage, we utilize weighted voting to determine the final retrieval outcomes. Our approach achieved 5th place in the monolingual track and 7th place in the crosslingual track. We release our system code at: \url{https://github.com/warmth27/SemEval2025_Task7}.

\end{abstract}

\section{Introduction}

SemEval-2025 Shared Task 7 focuses on the retrieval of monolingual and crosslingual fact-checked claims, aiming to tackle the global challenge of misinformation spread \cite{semeval2025task7}.

We engaged in two tracks of the SemEval-2025 Shared Task 7: monolingual and crosslingual. The monolingual track demands methods capable of retrieving the relationship between social media posts and fact-checked claims within the same linguistic environment. This task presents challenges such as noise arising from the large volume of data and difficulties related to the imbalance of language resources \cite{xu2024survey}. The crosslingual track requires methods that can retrieve fact-checked claims related to social media posts regardless of whether the language of the post matches the language of the related fact-checked claim. The primary challenge in crosslingual retrieval lies in translation inconsistencies, particulaedrly for low-resource languages \cite{qi2023cross, magueresse2020low}. The absence of high-quality translation tools exacerbates the complexity of achieving accurate crosslingual semantic alignment.

To tackle the aforementioned challenges, we propose a three-stage retrieval framework. Initially, we evaluate and employ several pre-trained language models for preliminary retrieval of candidate results \cite{gao2024multilingual, huang2024survey}, thereby mitigating the noise caused by the large data volume and alleviating the adverse effects of language resource imbalance. Subsequently, a re-ranking model is applied to refine the ranking of the candidate results, enhancing the position of fact-checked claims most relevant to the social media posts. For the crosslingual retrieval task, we utilize machine-translated data for preliminary retrieval, followed by ranking the results using a re-ranking model fine-tuned with English data. Finally, a weighted voting strategy is employed to combine the outputs from multiple re-ranking models, further enhancing the system’s accuracy.

Our approach achieved 5th place in the monolingual track and 7th place in the crosslingual track, thereby validating its effectiveness and feasibility in addressing the aforementioned challenges.

\section{System Description}

\begin{figure*}[t]
  \includegraphics[width=\textwidth]{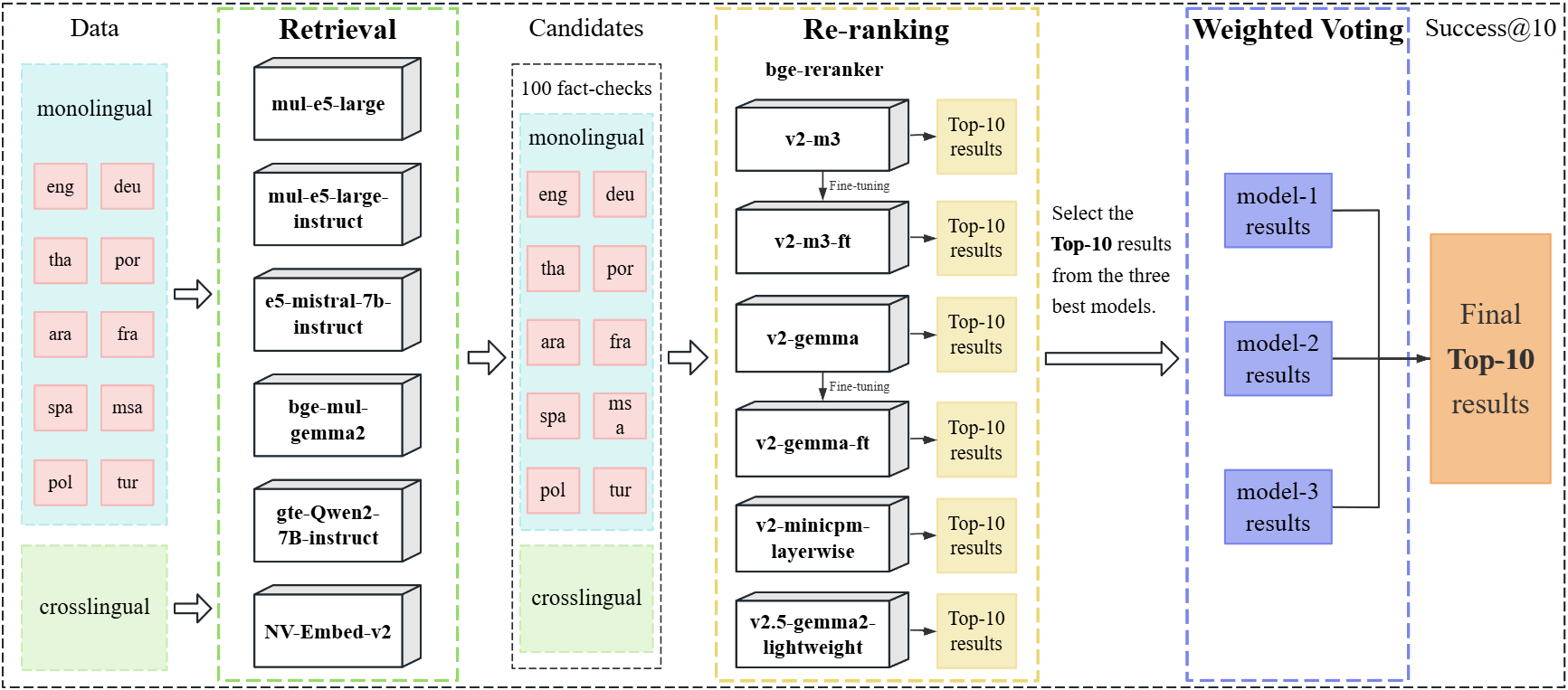}
  \caption{Illustration of the overall workflow in this paper.}
  \label{fig:1}
\end{figure*}

Our approach utilizes a three-stage retrieval framework: Retrieval stage, Re-ranking stage, and Weighted Voting stage. This staged design excels at balancing retrieval efficiency and accuracy, making it particularly suitable for handling large-scale datasets. By generating candidate results during the initial retrieval stage, fine-tuning them during the re-ranking phase, and finally aggregating predictions from multiple models in the weighted voting stage, we are able to obtain the final solution. The detailed process is shown in Figure \ref{fig:1}.

\subsection{Retrieval Stage}

The retrieval model calculates the similarity between the query and the documents, ranking them from most to least relevant based on their similarity. This aids in filtering out a subset of candidate data from a large pool of fact-checked claims, thereby reducing noise. Accordingly, we compare several pre-trained retrieval models and employ the top-performing models in each language to retrieve candidate fact-checked claims.

The key advantage of this strategy lies in significantly reducing the computational complexity of the subsequent re-ranking phase, effectively minimizing the noise caused by the large volume of candidate fact-checked claims by pre-generating the results. The choice of the number of candidate results ensures breadth while avoiding irrelevant information, providing sufficient diversity and selection space for the re-ranking phase, thus ensuring that the final output maintains high relevance.

\subsection{Re-ranking Stage}

The re-ranking model can make a more refined evaluation of the results of the initial retrieval stage, put the most relevant claims at the front, and further improve the accuracy of retrieval. Therefore, we select a series of re-ranking models and fine-tuning them using the data from the training set, training a set of re-ranking models for each language \cite{jiang2023team}. Additionally, we combine a re-ranking model with larger parameters to re-rank the 100 candidate results generated in the initial retrieval stage. Through more precise evaluation, we improve the ranking of the most relevant fact-checked claims related to the input social media posts, thereby optimizing the retrieval results.

\subsection{Weighted Voting Stage}

The weighted voting strategy combines the strengths of different models through weighted fusion, minimizing the potential biases and errors inherent in individual models \cite{wang2023ensemble}. Therefore, we adopt the weighted voting strategy to integrate the predictions of the re-ranking models. The weight of each re-ranking model is assigned based on its performance on the validation set, with better-performing models receiving higher weights. This ensemble method leverages the strengths of multiple models, synthesizing their predictions to produce more accurate top 10 (Top-10) results.
For the crosslingual task, we apply the same strategy.

\section{Experimental Setup}

\subsection{Data}

Following the guidelines, we partitioned the original dataset into both monolingual and crosslingual train and development (dev) datasets \cite{pikuliak2023multilingual}. For the monolingual data, we further categorize it by language, producing separate sub-train and sub-dev datasets for eight languages. The data statistics are presented in Appendix \ref{sec:tab1}



\subsection{Preprocessing}

In the data preprocessing phase, we transform the CSV file into JSON format, utilize regular expressions to extract key fields containing the original text and its translation, such as "ocr", "text", "title", and "claim", and separate them into original text, translated text, and language identifiers. Missing values (NaN) are consistently marked as null, and all text data is encoded in UTF-8 format.

\subsection{Evaluation Metrics}

The task employs Success@10 (S@10) as the primary evaluation metric. Specifically, when multiple correct fact-checked claims are present, the task is deemed successful if any one of the correct results appears on the Top-10 list, allowing the social media post to receive a score.

\section{Experiments and Results}



\subsection{Retrieval}

\begin{table*}[]
\centering
\resizebox{\textwidth}{!}{%
\begin{tabular}{l|l|ccccccccc}
\hline
\textbf{Model}                                   & \textbf{Plan}   & \textbf{Avg} & \multicolumn{1}{l}{\textbf{eng}} & \multicolumn{1}{l}{\textbf{spa}} & \multicolumn{1}{l}{\textbf{deu}} & \multicolumn{1}{l}{\textbf{por}} & \multicolumn{1}{l}{\textbf{fra}} & \multicolumn{1}{l}{\textbf{ara}} & \multicolumn{1}{l}{\textbf{msa}} & \multicolumn{1}{l}{\textbf{tha}} \\ \hline
\multirow{3}{*}{mul-e5-large}           & O      & 82.18     & 78.03                   & 81.30                   & 79.51                   & 80.46                   & 84.04                   & 80.76                   & 78.09                   & 95.23                   \\
                                        & T      & 82.11     & 78.87                   & 85.85                   & 72.28                   & 78.80                   & 85.10                   & 80.76                   & 80.00                   & 95.23                   \\
                                        & O, T   & 84.52     & 74.89                   & 86.99                   & 79.51                   & 83.77                   & 85.63                   & 85.89                   & 86.66                   & 92.85                   \\ \hline
\multirow{4}{*}{mul-e5-large-instruct}  & O      & 83.14     & 79.07                   & 86.99                   & 74.69                   & 76.49                   & 86.17                   & 80.76                   & 85.71                   & 95.23                   \\
                                        & T      & 83.53     & 78.03                   & 87.31                   & 68.67                   & 80.79                   & 85.63                   & 85.89                   & 86.66                   & 95.23                   \\
                                        & O, T   & 84.16     & 79.49                   & 86.50                   & 79.51                   & 77.81                   & 88.82                   & 85.89                   & 80.00                   & 95.23                   \\
                                        & O, T, V & 83.22     & 77.19                   & 85.20                   & 66.26                   & 81.78                   & 85.63                   & \textbf{89.74}                   & 84.76                   & 95.23                   \\ \hline
\multirow{4}{*}{e5-mistral-7b-instruct} & O      & 80.96     & 81.79                   & 84.06                   & 68.67                   & 78.80                   & 86.70                   & 74.35                   & 78.09                   & 95.23                   \\
                                        & T      & 85.20     & 82.00                   & 86.99                   & 75.90                   & 85.43                   & 88.29                   & 82.05                   & 85.71                   & 95.23                   \\
                                        & O, T   & 87.79     & 84.72                   & \textbf{91.21}                   & \textbf{80.72}                   & \textbf{88.41}                   & 92.55                   & 79.48                   & 87.61                   & \textbf{97.61}                   \\
                                        & O, T, V & \textbf{87.90 }    & \textbf{85.98}                   & 89.91                   & \textbf{80.72}                   & 87.41                   & \textbf{93.61}                   & 84.61                   & 85.71                   & 95.23                   \\ \hline
bge-mul-gemma2                          & O, T   & 87.71     & 81.38                   & 91.05                   & 79.51                   & 83.11                   & 90.42                   & 87.17                   & \textbf{91.42}                   & \textbf{97.61}                   \\ \hline
gte-Qwen2-7B-instruct                   & O, T   & 84.43     & 81.38                   & 88.78                   & 75.90                   & 80.46                   & 86.17                   & 79.48                   & 85.71                   & \textbf{97.61}                   \\ \hline
NV-Embed-v2                             & O, T   & 85.94     & 82.42                   & 87.47                   & 78.31                   & 81.45                   & 91.48                   & 85.89                   & 87.61                   & 92.85                   \\ \hline
\end{tabular}%
}
\caption{The Success@10 (S@10) scores (\%) for the monolingual track, where O uses original text, T uses translation, O,T combines both, and O,T,V includes the verdict field. \textbf{Bold} highlights the best score.}
\label{tab:2}
\end{table*}

\paragraph{Monolingual}

During the development phase, we conduct systematic experiments to confirm the effectiveness of the multilingual feature combination strategy. As shown in Table \ref{tab:2}, the combination of the original text (O) and the machine translated text (T) improves the S@10 score of the model, which means that the combination of the original text and the translated text (O, T) can improve the performance of the model on monolingual retrieval tasks \cite{muennighoff2022crosslingual}. The introduction of the translated text can help the model better understand the content of the original text, thereby improving the retrieval accuracy, especially when dealing with complex or ambiguous expressions.

Building upon the combined input of the original text and the translation, we further incorporate the "verdicts" field (V). The experimental results reveal that for several languages (Spanish, Portuguese, Malay, and Thai), the scores decrease by about 2\% after including the "verdicts" field. However, the scores for English, French, and Arabic improve after incorporating the "verdicts" field, with Arabic showing an increase of approximately 5\%. This suggests that in specific languages, the "verdicts" field can offer valuable supplementary information, aiding in the enhancement of retrieval accuracy.

Simultaneously, we compare the performance of several retrieval models, including mul-e5-large \cite{wang2024multilingual}, mul-e5-large-instruct, e5-mistral-7b-instruct \cite{wang2023improving}, bge-mul-gemma2 \cite{bge_embedding}, NV-Embed-v2 \cite{lee2024nv}, and gte-Qwen2-7B-instruct\footnote{\url{https://huggingface.co/models}} \cite{li2023towards}. As shown in Table \ref{tab:2}, the experimental results demonstrate that e5-mistral-7b-instruct achieves the best performance in eng (85.98\%), spa (91.21\%), deu (80.72\%), por (88.41\%), fra (93.61\%) and tha (97.61\%), while bge-mul-gemma2 performs better in msa (91.42\%) and mul-e5-large-instruct achieves the highest score in ara (89.74\%). This indicates that different retrieval models exhibit specific advantages or limitations depending on the language.

\paragraph{Crosslingual}

In the crosslingual task, we compare the performance of multiple retrieval models. The results shown in Table \ref{tab:3} demonstrate that the effect of using pure translation input is better than mixed input (combining original and translated text) and pure original text input. Among them, e5-mistral-7b-instruct performs the best, with S@10 reaching 72.64\%. In crosslingual retrieval, the semantic expression of the translation is more consistent and accurate,  thereby improving retrieval performance. On the contrary, combining the original and translated text or using pure original text input may introduce noise due to language differences, resulting in reduced retrieval performance. This further highlights the key role of translation consistency in crosslingual semantic matching.

\begin{table}[h]
\centering
\setlength{\tabcolsep}{13pt}   
\renewcommand{\arraystretch}{1.0}
\begin{tabular}{l|l|c}
\hline
\textbf{Model}                                  & \textbf{Plan} & \textbf{Avg} \\ \hline
\multirow{3}{*}{mul-e5-large}          & O    & 52.89      \\
                                       & T    & 58.51      \\
                                       & O, T & 46.92      \\ \hline
\multirow{3}{*}{mul-e5-large-instruct} & O    & 57.78      \\
                                       & T    & 62.86      \\
                                       & O, T & 58.87      \\ \hline
e5-mistral-7b-instruct                 & O, T & \textbf{72.64}      \\ \hline
bge-mul-gemma2                         & O, T & 71.37      \\ \hline
\end{tabular}
\caption{The Success@10 (S@10) scores (\%) of the crosslingual results. \textbf{Bold} indicates the best S@10.}
\label{tab:3}
\end{table}

\begin{table*}[]
\centering
\renewcommand{\arraystretch}{1.0}
\begin{tabular}{l|ccccccccc}
\hline
\textbf{Model}                  & \textbf{Avg} & \multicolumn{1}{l}{\textbf{eng}} & \multicolumn{1}{l}{\textbf{spa}} & \multicolumn{1}{l}{\textbf{deu}} & \multicolumn{1}{l}{\textbf{por}} & \multicolumn{1}{l}{\textbf{fra}} & \multicolumn{1}{l}{\textbf{ara}} & \multicolumn{1}{l}{\textbf{msa}} & \multicolumn{1}{l}{\textbf{tha}} \\ \hline
v2-m3                  & 72.83     & 72.17                   & 72.52                   & 65.06                   & 68.87                   & 79.78                   & 79.48                   & 59.04                   & 85.71                   \\
v2-m3-ft               & 92.57     & 89.33                   & 93.82                   & 89.15                   & \textbf{92.71}                   & \textbf{94.68}                   & \textbf{92.30}                    & 93.33                   & 95.23                   \\
v2-gemma               & 91.56     & 89.12                   & 93.17                   & 89.15                   & 88.74                   & 91.48                   & \textbf{92.30}                    & 88.57                   & \textbf{100.0}                   \\
v2-gemma-ft            & \textbf{93.73}     & \textbf{89.74}                   & \textbf{94.95}                   & \textbf{93.97}                   & \textbf{92.71}                   & 93.61                   & 91.02                   & \textbf{96.19}                   & 97.61                   \\
v2.5-gemma2-lightweight & 92.74     & 89.53                   & 93.82                   & 90.36                   & 92.05                   & 93.08                   & 89.74                   & 93.33                   & \textbf{100.0}                   \\
v2-minicpm-layerwise   & 87.25     & 86.82                   & 92.19                   & 80.72                   & 89.07                   & 90.95                   & 85.89                   & 86.66                   & 85.71                   \\ \hline
\textbf{Voting}                 & 95.14     & 91.21                   & 95.44                   & 93.97                   & 94.03                   & 95.74                   & 93.58                   & 97.14                   & 100.0                   \\ \hline
\end{tabular}
\caption{Results of the re-ranking model for the monolingual track. The -ft indicates the model fine-tuned on this model. \textbf{Bold} indicates the S@10 score (\%) of the best re-ranking model. \textbf{Voting} is the result of the third-stage weighted voting.}
\label{tab:4}
\end{table*}

\subsection{Re-ranking}

We utilize the models that performe well during the retrieval phase to extract 100 fact-checked claims as candidate data from the sub-dev sets of each language. Subsequently, we employ re-ranking models from the BAAI/bge-reranker\footnote{\url{https://huggingface.co/BAAI}} series to reorder the candidate data and derive the Top-10 as the re-ranking results.

\paragraph{Monolingual}

The experimental results in Table \ref{tab:4} demonstrate that v2.5-gemma2-lightweight outperforms the other models, achieving an S@10 score of 92.74\%. In comparison, v2-minicpm-layerwise and v2-m3 \cite{bge-m3} exhibit weaker performance. This outcome can be attributed to the disparities in model parameters. v2.5-gemma2-lightweight benefits from a larger parameter size and enhanced learning capability, enabling it to capture semantic information more efficiently. Conversely, v2-minicpm-layerwise and v2-m3 have relatively smaller parameter sizes, which hinder their ability to handle complex retrieval tasks, likely contributing to their suboptimal performance.

We experiment with applying the rerank model directly to reorder Arabic (ara) data, resulting in an S@10 score of 85.89\% . In comparison to using the retrieval model alone, the rerank model does not demonstrate a clear advantage and introduces additional computational costs. This implies that relying exclusively on the rerank model does not fully utilize the system's overall performance, and combining the retrieval model with the rerank model is clearly the more effective strategy.

Additionally, we conduct fine-tuning on the complete training set for the v2-m3 and v2-gemma models. The experimental results reveal that the fine-tuned models demonstrate a significant improvement, with gains of 19.74\% and 2.17\% over the original models, respectively. Remarkably, the fine-tuned v2-gemma outperforms the larger parameter model v2.5-gemma2-lightweight, which provides strong evidence of the effectiveness of fine-tuning the rerank model.

\paragraph{Crosslingual}

In crosslingual, we fine-tuning the v2-m3 and v2-gemma models using the translation data in the crosslingual training set and compare them with v2-gemma and v2.5-gemma2-lightweight \cite{cui2025multilingual}. Considering that the performance is better when only English translation data is used in crosslingual tasks, we also add a comparison with two models (v2-m3 and v2-gemma) fine-tuned on the English monolingual training set. The experimental results show that the v2.5-gemma2-lightweight model performs best with an S@10 of 80.25\%, while the model fine-tuned on crosslingual data performs worse than the model fine-tuned on English monolingual data. We speculate that this difference may be attributed to the quality of the translations in the crosslingual training set, which may affect the language representation learned by the model, resulting in poor crosslingual matching performance.

\begin{table}[h]
\centering
\setlength{\tabcolsep}{20pt}
\renewcommand{\arraystretch}{1.0}
\begin{tabular}{l|c}
\hline
\textbf{Model}                  & \textbf{Avg} \\ \hline
v2-m3-ft(cross)        & 76.44      \\
v2-m3-ft(eng)          & 79.16      \\
v2-gemma               & 75.72      \\
v2-gemma-ft(cross)     & 78.62      \\
v2-gemma-ft(eng)       & 79.34      \\
v2.5-gemma2-lightweight & \textbf{80.25}      \\ \hline
\textbf{Voting}                 & 84.05      \\ \hline
\end{tabular}
\caption{The results of the re-ranking model in the crosslingual track. The (cross) indicates the model fine-tuned using the crosslingual training set data, and the (eng) indicates the model fine-tuned utilizing the English monolingual training set data. \textbf{Bold} indicates the S@10 score (\%) of the best re-ranking model. \textbf{Voting} is the result of the third-stage weighted voting.}
\label{tab:5}
\end{table}

\subsection{Weighted Voting}

Finally, we employ a weighted voting ensemble strategy to integrate the results of the monolingual and cross-lingual re-ranking models in Table \ref{tab:4} and Table \ref{tab:5}. Experimental results demonstrate that the S@10 scores after integration are equal to or exceed those of the individual re-ranking models in all languages. The average of monolingual S@10 is 1.41\% higher than that of the highest re-ranking model, and the crosslingual S@10 is 3.8\% higher than that of the highest re-ranking model.

\subsection{Evaluation on the Test Set}

\paragraph{Test Data Augmentation}

For the newly introduced Polish (pol) and Turkish (tur) in the test set, we translate the original text from training set data in other languages into pol for data augmentation and to fine-tune the re-ranking model \cite{huang20241+, xu2024team}. The results reveal that the S@10 score of the re-ranking model v2-m3, fine-tuned using the data augmentation method on Polish, is 83.2\%, while the v2-gemma model achieved a score of 85.4\%. Both scores are lower than the 88.6\% score obtained by directly using the re-ranking model with a larger parameter size.

Furthermore, for Portuguese (por), which exhibited relatively low scores during the test phase, we aimed to augment the data by translating training data from other languages into Portuguese \cite{hangya2022improving}. The experiment showed that models fine-tuned with the augmented training data achieved an S@10 score of 87.2\% for Portuguese, representing a 1.4\% decrease compared to the previous models and falling short of the anticipated improvement.

This suggests that while translated data can enhance the dataset's diversity, the translation process may introduce semantic distortions and information loss. The meaning and context of the original text may not be entirely preserved, leading to issues like translation inconsistency and data mismatch, which prevent the model from benefiting from the augmented training data.

\paragraph{Official Test Results}

In the final test phase, based on the official evaluation metric S@10, our approach achieved the highest score of 93.64\% in the monolingual track, securing the 5th place (5/28). In the crosslingual track, our best score was 79.25\%, ranking 7th (7/29), further validating the effectiveness of our method. The scores of each language are shown in Table \ref{tab:6}.

\begin{table}[h]
\centering
\resizebox{\columnwidth}{!}{%
\begin{tabular}{cccccc}
\hline
\textbf{Mono\_Avg}               & eng                       & fra                       & deu                       & por                       & \multicolumn{1}{c}{spa} \\ \hline
93.65                   & \multicolumn{1}{c}{89.40} & \multicolumn{1}{c}{95.00} & \multicolumn{1}{c}{90.20} & \multicolumn{1}{c}{89.00} & 94.80                   \\ \hline
\hline
\multicolumn{1}{c}{tha} & msa                       & ara                       & tur                       & pol                       & \textbf{Cross\_Avg}              \\ \hline
99.45                   & \multicolumn{1}{c}{100.0} & \multicolumn{1}{c}{97.0}  & \multicolumn{1}{c}{93.00} & \multicolumn{1}{c}{88.60} & 79.25                   \\ \hline
\end{tabular}%
}
\caption{Final S@10 score (\%) on the official test set}
\label{tab:6}
\end{table}

\section{Conclusion and Limitation}

This paper introduces a monolingual and crosslingual fact-checked claim retrieval method utilizing a three-stage retrieval framework. By integrating retrieval models, re-ranking models, and weighted voting, we effectively address challenges such as data noise and imbalanced language resources. Our findings suggest that employing a mixed input strategy markedly enhances retrieval performance, while fine-tuning further optimizes re-ranking efficacy. Our method achieved 5th place in the monolingual track and 7th place in the crosslingual track.

We acknowledge that our method has limitations in terms of translation consistency and quality. Future work will focus on enhancing translation quality and refine model fine-tuning strategies to overcome these challenges.

\section*{Acknowledgments}

This work is funded by the Natural Science Foundation of Shandong Province under grant ZR2023QF151 and the Natural Science Foundation of China under grant 12303103.

\bibliography{custom}



\newpage
\appendix
\setcounter{table}{0}
\setcounter{figure}{0}
\renewcommand{\thetable}{A\arabic{table}}
\twocolumn
\section{Appendix}
\label{sec:tab1}

\begin{table}[htbp]
\centering
\renewcommand{\arraystretch}{1.0}
\setlength{\tabcolsep}{4.2pt} 
\begin{tabular}{l|rrrrr}
\hline
\multicolumn{1}{c|}{\multirow{2}{*}{\textbf{Mono}}} & \multicolumn{3}{c}{\textbf{Post}}    & \multicolumn{2}{c}{\textbf{Fact\_check}} \\ \cline{2-6} 
\multicolumn{1}{c|}{} & \multicolumn{1}{c}{train} & \multicolumn{1}{c}{dev} & \multicolumn{1}{c}{test} & \multicolumn{1}{c}{train, dev}     & \multicolumn{1}{c}{test}    \\ \hline
eng                                              & 4,351         & 478        & 500        & 85,734             & 145,287          \\
spa                                              & 5,628         & 615        & 500        & 14,082             & 25,440           \\
deu                                              & 667          & 83         & 500         & 4,996             & 7,485            \\
por                                              & 2,571         & 302        & 500        & 21,569             & 32,598           \\
fra                                              & 1,596         & 188        & 500         & 4,355             & 6,316            \\
ara                                              & 676          & 78         & 500        & 14,201             & 21,153           \\
msa                                              & 1,062         & 105        & 93         & 8,424              & 686             \\
tha                                              & 465          & 42         & 183          & 382             & 583             \\
pol                                              & -            & -          & 500            & -             & 8,796            \\
tur                                              & -            & -          & 500            & -             & 12,536           \\ \hline
\textbf{Cross}                                     & 4,972         & 552        & 4,000       & 153,743            & 272,447          \\ \hline
\end{tabular}
\caption{Statistics on monolingual and crosslingual tracks. These languages are: English (eng), Spanish (spa), German (deu), Portuguese (por), French (fra), Arabic (ara), Malay (msa), Thai (tha), Polish (pol) and Turkish (tur).}
\label{tab:1}
\end{table}

\end{document}